\documentclass[journal]{IEEEtran}
\usepackage{graphicx}
\usepackage[numbers]{natbib}
\usepackage{amsmath,amssymb,amsfonts}
\usepackage{multirow}
\usepackage{threeparttable}
\usepackage{hyperref}
\usepackage{caption}
\usepackage{subfig}
\usepackage{color}
\usepackage[labelfont=bf, justification=justified]{caption}
\ifCLASSINFOpdf
\else
\fi

\begin{document}
%
\title{Perception Consistency Ultrasound Image Super-resolution via Self-supervised CycleGAN}
%
%
%

\author{Heng~Liu,  
		Jianyong~Liu,
		Tao~Tao,
		Shudong~Hou and
		Jungong~Han

\thanks{H. Liu, J. Liu, T. Tao and S. Hou are with School of Computer Science and Technology, Anhui University of Technology, Ma'anshan 243032, China. E-mail: hengliusky@aliyun.com.}
\thanks{J. Han is with Aberystwyth University, SY23 3FL, UK. E-mail: jungonghan77@gmail.com.}}

\maketitle

\begin{abstract}
Due to the limitations of sensors, the transmission medium and the intrinsic properties of ultrasound, the quality of ultrasound imaging is always not ideal, especially its low spatial resolution. To remedy this situation, deep learning networks have been recently developed for ultrasound image super-resolution (SR) because of the powerful approximation capability. However, most current supervised SR methods are not suitable for ultrasound medical images because the medical image samples are always rare, and usually, there are no low-resolution (LR) and high-resolution (HR) training pairs in reality. In this work, based on self-supervision and cycle generative adversarial network (CycleGAN), we propose a new perception consistency ultrasound image super-resolution (SR) method, which only requires the LR ultrasound data and can ensure the re-degenerated image of the generated SR one to be consistent with the original LR image, and vice versa. We first generate the HR fathers and the LR sons of the test ultrasound LR image through image enhancement, and then make full use of the cycle loss of LR-SR-LR and HR-LR-SR and the adversarial characteristics of the discriminator to promote the generator to produce better perceptually consistent SR results. The evaluation of PSNR/IFC/SSIM, inference efficiency and visual effects under the benchmark CCA-US and CCA-US datasets illustrate our proposed approach is effective and superior to other state-of-the-art methods.
\end{abstract}


%
\IEEEpeerreviewmaketitle

\section{Introduction}
%
%
%
%
\IEEEPARstart{M}{edical} imaging is an effective and widely used diagnosis tool in this modern medical industry, which commonly includes ultrasound imaging, magnetic resonance imaging (MRI), X-ray and computed tomography (CT). Among them, ultrasound imaging has the characteristics of low cost, non-radiation and continuous dynamic recording, which is superior to others. In the actual ultrasound imaging diagnosis, doctors usually judge whether there is a lesion by observing the shape, the blood flow degree, and the contour smoothness of the interest region in the ultrasound images. This indicates that the high resolution of ultrasound images is conducive to improving the accuracy of medical diagnosis. Actually, due to the limitation of acoustic diffraction in medical equipment, it is hard to obtain HR ultrasonic data. Thus, in terms of improving the resolution of ultrasound data, image super-resolution turns out to be a feasible approach, which is of great importance for visual perception based medical clinical diagnosis \cite{hudson2015dynamic,morin2013semi}. 

In the last couple of years, deep learning network has been applied to a variety of medical image processing tasks, including CT image segmentation \cite{skourt2018lung}, MRI image deblurring \cite{lim2020deblurring} and ultrasound image SR \cite{choi2018deep,lu2018unsupervised}. Umehara \emph{et al.} \cite{umehara2018application} in the first time applied the deep neural network to medial images. They improved the resolution of CT images with the pioneer image SR model - SRCNN \cite{dong2016image}). Recent works on bio-medical image segmentation and ultrasound image SR \cite{Olaf2015Unet,van2019deep} utilized the classical  ’U-net’ structure to develop the task-specific deep models. Since there is no fully connected layer, the overall structure of U-Net is made up of many convolution and deconvolution layers. Here convolution layer plays a role of encoder while deconvolution layer acts as a decoder.  Actually, the pooling operations and the single-scale structure in such U-net model may not be able to make full use of the multi-level image details and the multi-scope context information. 

A recent work \cite{kim2016accurate} suggested that better SR results can be acquired through a deeper and wider network with good generalization performance. In practice, this principle may not be always applicable to medical imaging field due to the fact that usually there are not numerous medical LR-HR sample pairs available for supervision training. Therefore, how to deal with the lack of supervision samples becomes one of the keys to improving the performance of medical image SR.

Different from CNNs, Ledig \emph{et al.} \cite{ledig2017photo} introduced the idea of adversarial learning for image generation to produce photo-realistic SR results, and form a new network structure, namely SRGAN (SR generative adversarial network). Also the SRGAN model had been applied by Choi \emph{et al.} \cite{choi2018deep} for high-speed ultrasound image SR. Moreover, Yochai \emph{et al.} in their recent work \cite{blau2018perception} found that although GANs can obtain better reconstruction effect, the visual perception quality and the distortion decreasing metric seem to be contradictory with each other. 

In fact, the aforementioned deep SR methods are all working in the way of supervised learning with numerous LR-HR samples pairs, and are not suitable for unsupervised or self-supervised scenario. Meanwhile, these methods don't consider the consistency from LR to SR and then back to LR again. Thus, in this work, motivated by zero-shot natural image SR (ZSSR) \cite{ZSSR} and CycleGAN \cite{zhu2017unpaired}, we present a novel self-supervised CycleGAN framework for ultrasound image SR, which is fully different from the structures of ZSSR \cite{ZSSR} and CycleGAN \cite{zhu2017unpaired}. In our approach, for LR to SR , we firstly construct deep multi-scale encoder-decoder \cite{liu2020exploring} to super-resolve the LR input. And then, for back to LR, we use a special designed CNN with random noise input to degenerate the generated SR one. While for HR to LR and then back to SR, these two structures just utilized are used again in reverse order.

Due to the cycle consistency structure, our proposed model greatly reduces the artifacts in SR results compared to ZSSR \cite{ZSSR}. Moreover, our model integrates multi-level feature loss when super-resolving ultrasound images to better balance the visual similarity to real data and the reconstruction accuracy. Numerous experimental comparisons under different ultrasound data sets are performed and the results show that the proposed approach can not only get good subjective visual effect but also obtain better objective quality evaluation metrics. 

Note that this work is a completely new development of our previous conference one \cite{liu2020exploring}. There are two obvious differences between them: the self-supervision learning mechanism is introduced to replace the previous supervised way; the CycleGAN structure with a richer variety of image losses including the cycle consistent loss is developed to replace the previous PatchGAN model. On the whole, our current work has made significant improvements on previous conference version and will get much better results than before.

To the best of our knowledge, there are few works to deal with the problem of deep SR for single ultrasound image, let alone exploring the self-supervision and cycle adversarial learning in the absence of LR-HR training pairs to realize accurate reconstruction with perception consistency. The contributions of this work can be summarized as follows:
\begin{itemize}
	\item By introducing the self-supervision mechanism with cycle adversarial learning, for the first time, we put forward a new self-supervised CycleGAN framework for single ultrasound image SR, which can lead to accurate reconstruction with perception consistency.
	\item Our proposed approach can adapt to ideal ultrasound images as well as non-ideal ones due to the self-supervision characteristics.
	\item We adopt both LR cycle loss and HR cycle loss with other multi-level image losses to jointly supervise the ultrasound image SR reconstruction. The experimental results indicate that the comprehensive loss can recover the multi-level and degradation consistent details of ultrasound images.
	\item We evaluate our approach on different public ultrasound datasets and provide the competitive results compared to other state-of-the-art methods. We also provide the ablation study on the proposed approach, which may be helpful for future further research on ultrasound image SR.
\end{itemize}

\section{Related Works}
\label{sec:RW}
\subsection{Natural Image SR}
Although image SR is a classic low-level vision task, it is still a research hot-spot in recent years, and many new methods have emerged, especially those based on deep learning. Since the advent of SRCNN - the first image SR deep network presented by Dong \emph{et al.} \cite{dong2016image}, many early deep SR models followed the process of feature extraction, nonlinear mapping and image reconstruction. However, such shallow neural networks hold the limited ability in obtaining multi-level features of the input images. With paying attention to that the edge prior is conducive to image SR, Liang \emph{et al.} \cite{liang2016incorporating} firstly utilized Sobel edges with LR images to train deep SR model. However, their  SR performance improvement is not obvious. Lately, based on the structure simulation on multiple resolution wavelet analysis, Liu \emph{et al.} \cite{liu2019single} proposed a multi-scale deep encoder-decoder model with the guidance of phase congruency edge map for single image SR and provided a convincing SR contrast effects. In addition, Wang \emph{et al.} \cite{wang2019multi} presented to form multi-memory residual block to progressively extract and retain inter-frame temporal correlations for video SR. Ma \emph{et al.} \cite{ma2020image} recently proposed dense discriminative network that is composed of several aggregation modules for image SR.

With applying adversarial learning strategy to improve the reconstruction quality, Ledig \emph{et al.} \cite{ledig2017photo} applied GAN's framework to present SRGAN for image SR. In the model, the generator utilizes several residual blocks for efficient SR reconstruction while the discriminator forces  the generator to produce the SR outputs close to the real HR labels. 

In addition, considering that the BN (batch normalization) operation may weaken the diversity of features, Lim \emph{et al.} \cite{lim2017enhanced} presented the so-called EDSR model by removing the BN layers in original deep residual blocks. They also made another adjustment to remove the ReLU layer after the sum of different paths so as to keep the path flexible.

Recently, Park \emph{et al.} \cite{park2018srfeat} presented a new GAN-like model - SRFeat, which holds two discriminators to not only distinguish the generated images but also the hierarchical features in the feature domain. This additional discrimination network can force the generator to pay attention on feature approximation while generating SR images. 

Completely different from above supervised methods, Shocher \emph{et al.} proposed a zero-shot image SR approach (ZSSR) \cite{ZSSR} which can work in unsupervised way. The ZSSR approach does not need the HR label data prepared in advance and can adapt to known as well as unknown imaging conditions theoretically. However, this method makes use of the pattern similarity of the image itself, and it is easy to produce the artifacts when applied to unnatural images such as medical ones.

\vspace{-2ex}
\subsection{Ultrasound Image SR}
Different from the vigorous development of natural image processing, medical image SR has not attracted enough attention. Recently, Zhao \emph{et al.} \cite{zhao2016single}  implemented ultrasound image SR by obtaining a $\ell_2$ norm regularization based analytical solution. Diamantis \emph{et al.} \cite{diamantis2018super} focused on axial imaging. They developed a location-based approach to convert SR axial imaging to ultrasound one and  recognized that the accuracy of ultrasonic axial imaging is closely related to the image-based location precision of single scattering.

Umehara \emph{et al.} \cite{umehara2018application} suggested the SRCNN approach might also be suitable to medical images, so they applied the method for chest CT image SR and the results supported their viewpoint. Moreover, similarly to ZSSR \cite{ZSSR}, Lu \emph{et al.} \cite{lu2018unsupervised} presented to exploit the multi-scale contextual features extracted from the test image itself to train an image-specific network and called this as unsupervised way, then utilized dilated convolution and residual learning to improve the convergence and accuracy.

In recently, U-Net \cite{Olaf2015Unet} deep network was applied by Van Sloun \emph{et al.} \cite{van2019deep} to super-resolve the vascular images based on high-density contrast-enhanced ultrasound data. In order to enhance details reconstruction in SR, Choi \emph{et al.} \cite{choi2018deep} slightly amended SRGAN \cite{ledig2017photo} model to enhance the transverse resolution of ultrasound images. Although the performance of adopting GAN is generally good, some recent study \cite{zhu2019make} have shown that the generated SR images can easily contain some unrealistic artificial details. This phenomenon has also been observed in our experiments (see Fig. \ref{fig5} and Fig. \ref{fig6} in this work). In addition, Liu \emph{et al.} \cite{Liu2019MedicalIS} presented to use dense connection with blended attention structure for MRI image SR. Although they  gave some quite good experimental results, their methods did not consider the image generation consistency of HR-to-LR and LR-to-HR.

\section{Methodology}
\subsection{Self-supervised ultrasound image SR}
Unlike other image processing tasks of low-level vision, image SR is to find a mapping function, which can map a LR image in LR image space onto a corresponding HR one in HR image space. Due to the different sources of various images, this mapping is usually complex and changeable. Therefore, whether or not the mapping relationship between the high resolution and the low one can be obtained accurately has a great impact on the SR performance. For natural images, this mapping can be gotten from a large number of pre-set LR-HR training sample pairs through supervised learning. But for ultrasound medical images, the situation is very different.

Ultrasound images usually come from clinical diagnosis, due to the privacy, it is difficult to obtain a great deal of training sample pairs for supervised learning. Even if such samples can be obtained, due to the different imaging conditions and acquisition scenes, it is difficult to find the accurate mapping relationship from ultrasound LR images to HR ones by supervised learning way.

However, due to the internal characteristics of ultrasound images, the changes of its edge and texture are relatively small compared with the natural image, and he content pattern has strong repeatability. Therefore, it is possible to exploit the relationship between the local region and the global image to construct training sample pairs and obtain the resolution mapping relationship at a specific down-sampling scale through self-supervised learning. Note that at this point, a general lightweight CNN network can meet the requirements. Actually, multi-scale analysis naturally has the excellent characteristics of capturing the relationship be tween the local region and the global image. Therefore, if we can build a multi-scale deep SR network, theoretically it will be more conducive to the performance improvement of this self-supervision learning method (will be described in detail in the following sections).

Our self-supervised ultrasound image SR approach can be described as follows: firstly, the test ultrasound image is made data enhancement and these enhanced images can be called ``HR fathers''; then these ``HR fathers'' are down-sampled at a specified reduction factor to obtain the "LR sons"; then a CycleGAN SR network is constructed, which utilizes multi-scale structure as the generator and considers the perception consistency from LR to HR and back to LR (which will be introduced in detail below); and then the LR-HR data pairs obtained before are used for the network training; finally, after the CycleGAN is well trained, the test ultrasound image is then sent to the generator as LR input to obtain its SR reconstruction result.

Note that above data enhancement operations on test ultrasound image include a series of down-sampling with different reduction factors, as well as 4 rotations ($0^{\circ}$, $90^{\circ}$, $180^{\circ}$, $270^{\circ}$) and their mirror reflections in the vertical and horizontal directions. In addition, for the purpose of robustness, we can also consider training several SR networks for certain intermediate down-sampling factors. The SR images generated by these networks and the corresponding down-scaled LR versions can also be added into the target training set as additional LR-HR example pairs.
\subsection{Multi-scale Generator}
\begin{figure}[t]
	\centering
	\includegraphics[width=3.2in, height=1.4in]{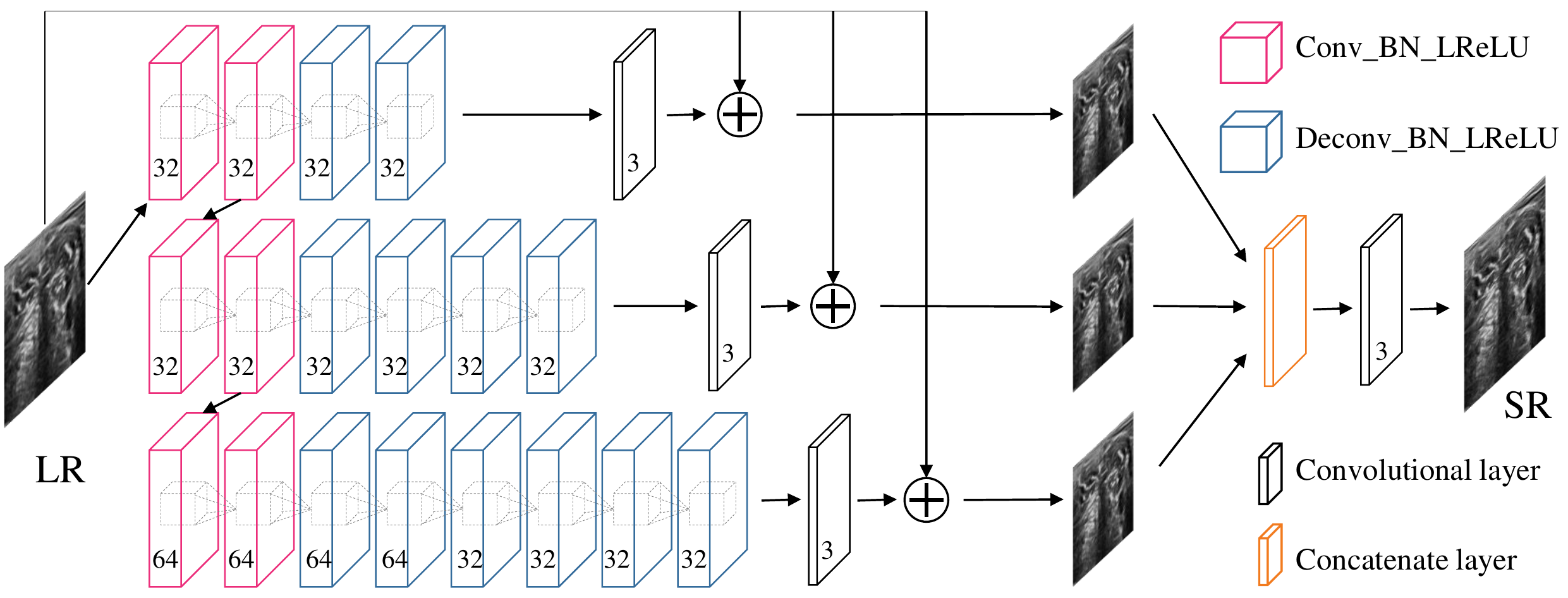}
	\caption{The structure of our multi-scale generator.}
	\vspace{-3ex}
	\label{fig1}
\end{figure}

Based on wavelet multi-resolution analysis (MRA) theory \cite{mallat1999wavelet} and motivated by the work \cite{liu2019single}, we can use deep structure to simulate wavelet multi-resolution analysis and construct a multi-scale deep network for ultrasound image SR. In order to adapt to any image size, our multi-scale model also adopts full convolution structure, which is fully composed of many encoders (convolution layers) and decoders (deconvolution layers). The detailed structure of our multi-scale generator is shown in Fig. \ref{fig1}. Note that this figure clearly demonstrates that the input LR image is considered to be the low frequency component of the multi-scale analysis of the HD image. Table \ref{table1} gives the detail parameters of our three-scale deep network. Obviously, the objective of multi-scale encoder-decoder learning is to find the optimized network parameters $\Theta_j$ of the network mapping function $F_j$ in every scale $j$ branch so that the final reconstruction can approximate the original HR image under certain measure (for example, $\ell_2$ norm). This may be formulated as: 
\begin{equation}
\tilde{\Theta}=\mathop{\arg\min_\Theta(||conv({\mathop{concat}\limits_{j}(\cdots,y+{F_j(y,\Theta_j)},\cdots))-f}||_2},
\label{eq1}
\end{equation}
where $f$ and $y$ are the HR image and the LR input, respectively. The symbol $j$ denotes a specific scale, the $concat(\cdot)$ formula means concatenation operation and the $conv(\cdot)$ represents the final output convolution operation in Fig. \ref{fig1}.
\begin{table}[h]
	\renewcommand{\arraystretch}{1.5}
	\vspace{-2ex}
	\caption{The specific parameters of three-scale generator}
	\label{table1}
	\centering
	\begin{threeparttable}[b]
		\begin{tabular}{c|c|c}
			\hline
			\bfseries scale1           & \bfseries scale2    & \bfseries scale3   \\
			\hline
			(conv3-32)$\times$2 & (conv3-32)$\times$2   & (conv3-32)$\times$2   \\
			& (conv3-32)$\times$2   & (conv3-32)$\times$2   \\
			&                       & (conv3-64)$\times$2   \\
			& 						& (deconv3-64)$\times$2 \\
			& (deconv3-32)$\times$2 & (deconv3-32)$\times$2	  \\
			(deconv3-32)$\times$2   & (deconv3-32)$\times$2 & (deconv3-32)$\times$2 \\
			\hline
		\end{tabular}
	\end{threeparttable}
	\vspace{-2ex}
\end{table}

In the multiple scales network, the LR image $I_{LR}$ is firstly input to three scales encoder-decoder streams to recover the image details at different scales. Since LR images can be treated as low-frequency components of HR ones (see Eq. (\ref{eq1}), the reconstructed images of different scales can be obtained by adding these detail images directly to LR input. Finally, the super-resolved ultrasound image $I_{SR}$ is obtained by concatenating and fusing the reconstruction images of three scales.

In fact, the multi-scale deep encoder-decoder structure acts as the generator of the CycleGAN based ultrasound image perception consistency SR framework, which will be described at length below. 
\subsection{CycleGAN based Perception Consistency  SR}
Different from traditional GAN \cite{goodfellow2014generative} that only contains one generator and one discriminator, CycleGAN \cite{zhu2017unpaired} employs two generators and discriminators to distinguish the generated images from real ones, equipping with the cycle consistency loss for reliable image generation.

Obviously, for medical image SR, the cycle consistency is particular significant because the redundant or artificial details introduced in image generation will seriously damage the accuracy of disease diagnosis. This fact is also a important motivation for us to use CycleGAN framework for ultrasound image SR. 

Since the original task of CycleGAN is image translation, it is easy to find a deal of natural images (paired or unpaired) for training. Whereas for ultrasound image SR, to obtain numerous paired LR and HR ultrasound images are quite difficult. Therefore, we not only need to build LR-to-HR generation model but also HR-to-LR one. Although the multi-scale deep encoder-decoder network mentioned above can be used as LR-to-HR generator, the HR-to-LR one still needs to be carefully designed and trained. 

\begin{figure}[t]
	\vspace{-1ex}
	\centering
	\includegraphics[width=3.3in, height=1.2in]{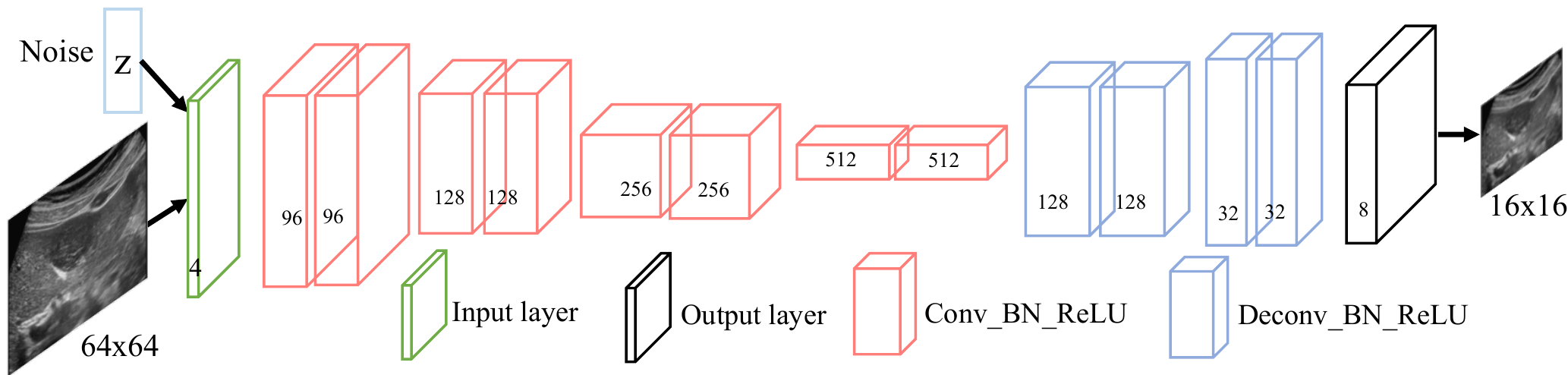}
	\caption{The detailed structure of our HR-to-LR ultrasound image generation network.}
	\vspace{-2ex}
	\label{fig2}
\end{figure}

Actually, as discussed in \cite{liu2020fast}, the HR-to-LR generation is just the complex image degradation process, which may involve multiple degeneration factors, such as noising, blurring and resolution decreasing. Fortunately, illuminated by the work \cite{bulat2018learn}, we introduce Gauss noise accompanied with LR image as input and construct a fully convolutional network (FCN) model to fulfill degrading high-resolution ultrasound image to LR one. The detail structure of our HR-to-LR ultrasound image generation network is shown in Fig. \ref{fig2}.
It should be noted that although the actual size of the output image of the HR-to-LR network is $1/4$ of the input image, for the convenience of calculating the HR consistency loss later, we will up-sample the output image to its $4$ times size.   

\begin{figure}[h]
	\vspace{-2ex}
	\centering
	\includegraphics[width=3.5in, height=1.2in]{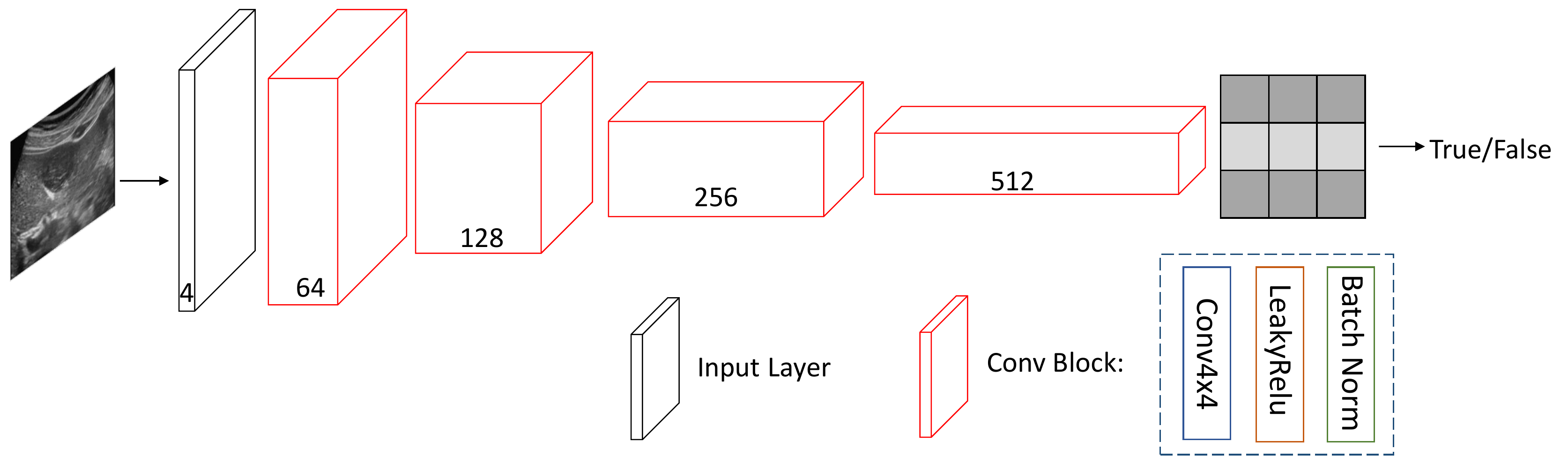}
	\caption{The detailed structure of the discriminator.}
	\vspace{-1ex}
	\label{fig3}
\end{figure}

Our perception consistency ultrasound image SR model contains two sets of GANs, each of which utilizes two generators (one is for LR and the other is for HR) and one patch discriminator. The two generators are composed of above multi-scale encoder-decoder and HR degradation network while the discriminator is mainly made with a input layer and four convolutional block, each block containing a convolutional layer, ReLU layer and batch normalization layer. The detail structure of the discriminator is shown in Fig. \ref{fig3}. The input to discriminator is the pair of the produced SR and the label HR or the pair of the generated LR and the label LR, all with size of $64\times64$.The output of the discriminator is an array $X$, where each $X_{ij}$ signifies whether the patch $ij$ in the image is real or fake.

\begin{figure*}[tb]
	\centering
	\includegraphics[width=6.4in, height=2.9in]{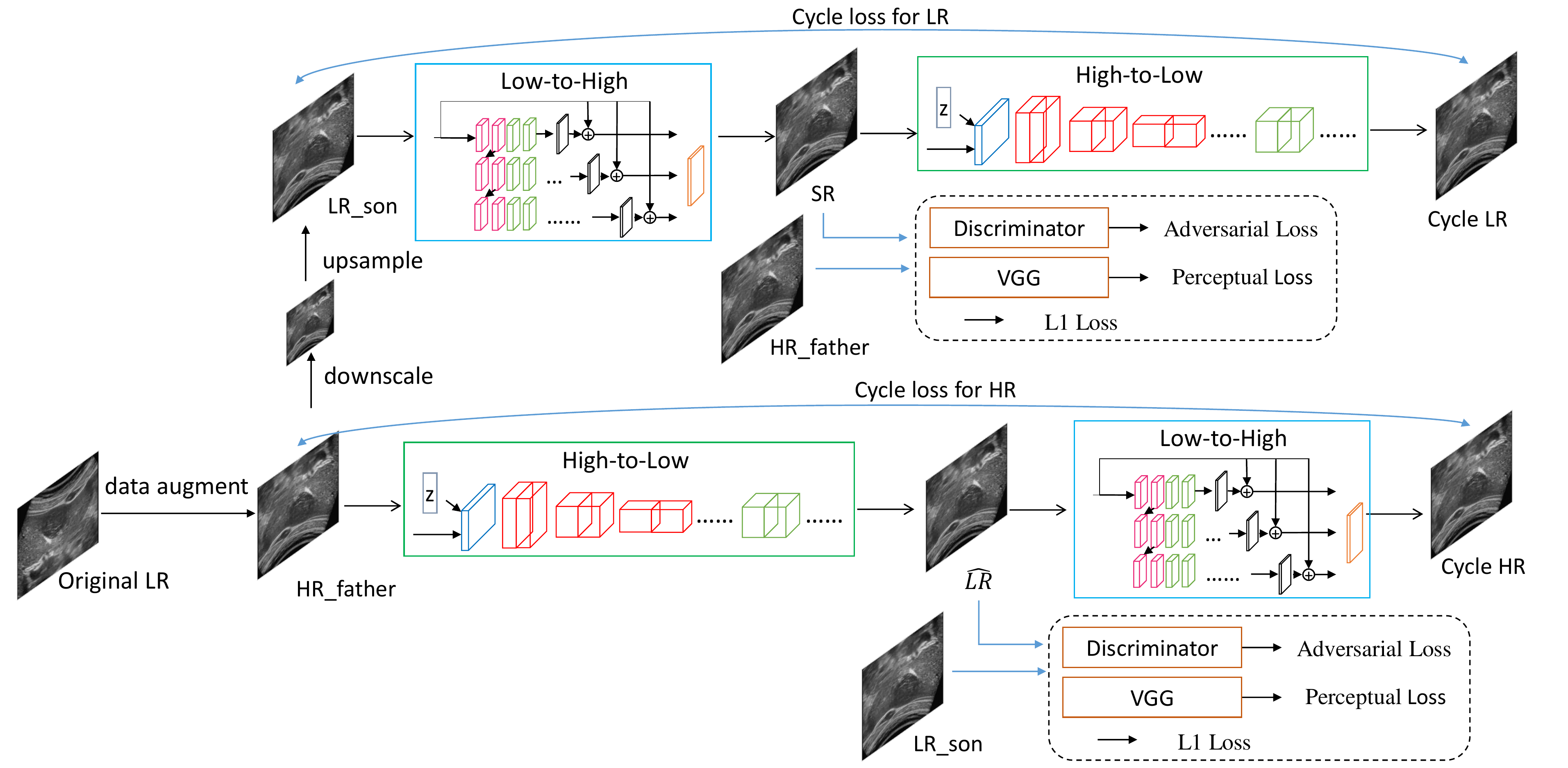}
	\caption{The proposed perception consistency ultrasound image SR model. The low-to-high generator (blue box) is the multi-scale encoder-decoder in Fig. \ref{fig1} and the high-to-low generator (green box) is the HR-to-LR degradation network in Fig. \ref{fig2}; the discriminator can be looked upon Fig. \ref{fig3} .}
	\vspace{-3ex}
	\label{fig4}
\end{figure*}

Our overall model can be looked upon as a CycleGAN framework, which includes two parts: one is LR cycle consistency GAN, the other is HR cycle consistency GAN. In addition, cycle consistency loss with multiple levels image measurement losses are introduced in the model. The architecture of our proposed model is illustrated in Fig. \ref{fig4}. In this figure, the detail structure of the low-to-high generator, the high-to-low generator and the discriminator can be refereed to Fig. \ref{fig1}, Fig. \ref{fig2} and Fig. \ref{fig3}, respectively.
\vspace{-2ex}
\subsection{Loss Function}
In order to ensure the perceptual consistency before and after ultrasound image generation, We firstly introduce the cycle losses for the generated cycle-HR and cycle-LR images, respectively. Since some recent works \cite{ledig2017photo,isola2017image} argued that using MSE loss in deep image generation training will incline to produce over-smooth results, we use $\ell_1$ loss instead of MSE ($\ell_2$) loss as a metric of the pixels difference between the generated one and the ground truth. Besides $\ell_1$ pixels proximity loss, we also incorporate other three levels loss functions to supervise SR or degradation to approximate the ground-truth one at multiple levels of details.

Given a set of LR and HR image pairs $\{x_i,y_i\}_{i=1}^N$ and assuming the low-to-high mapping function is $G:LR$ $\rightarrow$$HR$ and the high-to-low one is $F:HR$$\rightarrow$$LR$, then the $\ell_1$ pixel-wise loss for both low-to-high and high-to-low mappings can be denoted as:
\begin{equation}
\mathcal{L}_{pixel}=\frac{1}{N}(\sum_{i=1}^N(||G(x_i)-y_i||_1+||F(y_i)-x_i||_1))
\label{eq4}
\end{equation}

Besides the pixel-wise loss, since the perceptual loss is more beneficial to retention image features, we also make use of the perceptual loss when acquiring super-resolved or degraded ultrasound images. Specifically, we utilize the feature extraction function $\phi(\cdot)$ to transform $y_i$ and $x_i$ into certain common feature space . Then the distance between the two features in such feature space can be easily calculated. Commonly, the perceptual (feature) loss can be expressed as:
\begin{equation}
\begin{split}
\mathcal{L}_{percp}=\frac{1}{N}(\sum_{i=1}^N(||\phi(G(x_i))-\phi(y_i)||_2\\+||\phi(F(y_i))-\phi(x_i)||_2)),
\label{eq5}
\end{split}
\end{equation}
where the mapping function $\phi(\cdot)$ used in practice is the output combination of the 12th convolution layers from VGG \cite{simonyan2014very} network. 

We also apply the adversarial loss \cite{goodfellow2014generative} to both low-to-high and high-to-low generation networks. For the low-to-high generator $G:LR$ $\rightarrow$$HR$ and its discriminator $D_{hr}$, the adversarial loss for the generator may be expressed as: 
\begin{equation}
\mathcal{L}_{g\_adv}=\frac{1}{N}\sum_{i=1}^N
-log(D_{hr}(G(x_i)))
\label{eq7}
\end{equation}

Similarly, the adversarial loss for high-to-low generator $F:HR$$\rightarrow$$LR$ and its discriminator $D_{lr}$ can also be easily calculated, denoted as $\mathcal{L}_{f\_adv}$. Therefore, the total adversarial loss for such two GANs' generation mapping can be written as:
\begin{equation}
\mathcal{L}_{adv}=\frac{1}{N}\sum_{i=1}^N
(-log(D_{hr}(G(x_i)))-log(D_{lr}(F(y_i))))
\label{eq8}
\end{equation}

Although the adversarial loss can force the distribution of the generated SR image approximates to the distribution of the target HR data, it is not enough to guarantee that the learned mapping function can map an individual input $x_i$ to an expected target output $y_i$. In view of this, we introduce the LR-to-HR-to-LR and HR-to-LR-to-HR cycle losses to assure perception consistency for accurate ultrasound image reconstruction. Thus, the total cycle consistency loss may be formulated as:
\begin{equation}
\mathcal{L}_{cyc}=\frac{1}{N}(\sum_{i=1}^N(||F(G(x_i))-x_i||_1+||G(F(y_i))-y_i||_1))
\label{eq9}
\end{equation}

Finally, the total loss of our overall model is the sum of all the above losses and can be described as:
\begin{equation}
\mathcal{L}_{total}=\alpha\mathcal{L}_{pixel}+\beta\mathcal{L}_{percp}+\gamma\mathcal{L}_{adv}+\eta\mathcal{L}_{cyc},
\label{eq10}
\end{equation}
where $\alpha$, $\beta$, $\gamma$, and $\eta$ are the weighting coefficients, which control the relative importance of these different losses. 

In Section \uppercase\expandafter{\romannumeral4}, we will do the ablation study of some losses to show that the cycle structure and the consistency loss play important role in arriving at high-quality SR results. 

\section{Experimental results and analysis}
\subsection{Datasets}
The two public available ultrasound image datasets: CCA-US\footnotemark[1] and US-CASE\footnotemark[2] are fully used in this work to perform the SR experiments and the comparisons. The CCA-US data is acquired from ten volunteers with different ages and body weights (mean ages: $27.5 \pm 3.5$ years; mean weight: $76.5 \pm 9.7$ kg) by Sonix OP ultrasound scanner, which totally includes 84 B-mode ultrasound images of common carotid artery (CCA). While the US-CASE one is a free ultrasound library offered by SonoSkills and Hitachi Medical Systems Europe, which contains 125 ultrasound images of liver, heart and mediastinum, etc. Moreover, the well-known PSNR [dB], IFC \cite{sheikh2005information}, and SSIM \cite{wang2004image} metrics are exploited to evaluate the objective quality of the super-resolved ultrasound images. Our code for this work can be found at \url{https://github.com/hengliusky/UltraSound_SSSR}. 
\footnotetext[1]{http://splab.cz/en/download/databaze/ultrasound}
\footnotetext[2]{http://www.ultrasoundcases.info/Cases-Home.aspx}

\vspace{-4ex}
\subsection{Training Details}
The Original LR input can be any ultrasound image from these two datasets mentioned above. As described in section \uppercase\expandafter{\romannumeral3}, we can obtain ``HR fathers'' and ``LR sons'' from one image itself. We follow the strategy of ZSSR\cite{ZSSR} that training with random augmented cropped image instead of full image. Specifically, we obtain fixed-size random crops from father-son pair.The cropped size is typically set to $64 \times 64$ pixels. 

During the training, we utilize the total loss described in Eq. \ref{eq10} with Adam optimizer, starting with a learning rate of 0.001. The weighting coefficients of the loss function $\alpha$, $\beta$, $\gamma$ and $\eta$ are empirically set with 5, 0.1, 5 and 0.3, respectively. We also adopt the learning rate adjustment policy of ZSSR \cite{ZSSR} to gradually reduce the learning rate of our deep model. We stop training when the learning rate reaches to 0.000001. In order to stabilize the training, we follow the strategy of the work \cite{shrivastava2017learning} to update the discriminator with the historical generated images to avoid model oscillation. Finally, we combine self-ensemble and back projection techniques to get corrected median image as final super-resolved image.

\vspace{-2ex}
\subsection{Experimental Comparisons and Analysis}
Different ultrasound image SR methods are made comparative evaluation by performing 4$\times$ SR experiments. Note that the codes and the data set of current most medical image SR methods are not released. For example, Choi \emph{et al.} \cite{choi2018deep} and Lu \emph{et al.} \cite{lu2018unsupervised} individually utilize the slightly changed SRGAN \cite{ledig2017photo} and the convolution network with residual connection for ultrasound image SR. But they do not release their code and the ultrasound dataset. Fortunately, many recent natural image SR approaches, including the same or the very similar methods by  Choi \emph{et al.} and Lu \emph{et al.}, such as SRCNN \cite{dong2016image}, SRCAN \cite{ledig2017photo}, EDSR \cite{lim2017enhanced} (convolution network with residual connection), SRFeat \cite{park2018srfeat} have been public available. Therefore, we believe that the comparison results can correctly reflect the ultrasound image SR performance of the corresponding methods. In addition, for fair play, we use the two public datasets - CCA-US and US-CASE for comparisons. 
\begin{table}[b]
	\renewcommand{\arraystretch}{1.5}
	\vspace{-2ex}
	\caption{A comparison of PSNR and IFC scores under the test ultrasound dataset from US-CASE and CCA-US. The bold numbers indicate the best results.}
	\label{table2}
	\renewcommand\tabcolsep{8.0pt}
	\centering
	\vspace{-2ex}
	\begin{threeparttable}[t]
		\begin{tabular}{l|l l l l}
			\hline
			\multirow{2}{*}{\textbf{DataSets}} & \multicolumn{2}{c}{US-CASE} & \multicolumn{2}{c}{CCA-US}				\\ \cline{2-5}
			& PSNR		& IFC		& PSNR		& IFC     \\
			\hline
			Bicubic                          & 20.9131	& 1.213 & 26.300 & 1.055 \\
			
			SRCNN\cite{dong2016image}                            & 20.673	& 0.972 & 25.636 & 1.009 \\
			
			SRGAN\cite{ledig2017photo}                            & 25.331	& 1.127 & 29.069 & 1.102 \\

			\hline
			Our proposed                              
			&  \textbf{30.404}& \textbf{2.670} & \textbf{34.900} & \textbf{2.317} \\
			\hline
		\end{tabular}
	\end{threeparttable}
	\vspace{-2ex}
\end{table}

\begin{figure*}[t]
	\captionsetup[subfigure]{justification=centering}
	\centering
	\vspace{-1ex}
	\subfloat[Ground Truth] {\includegraphics[height=1.0in,width=1.3in]{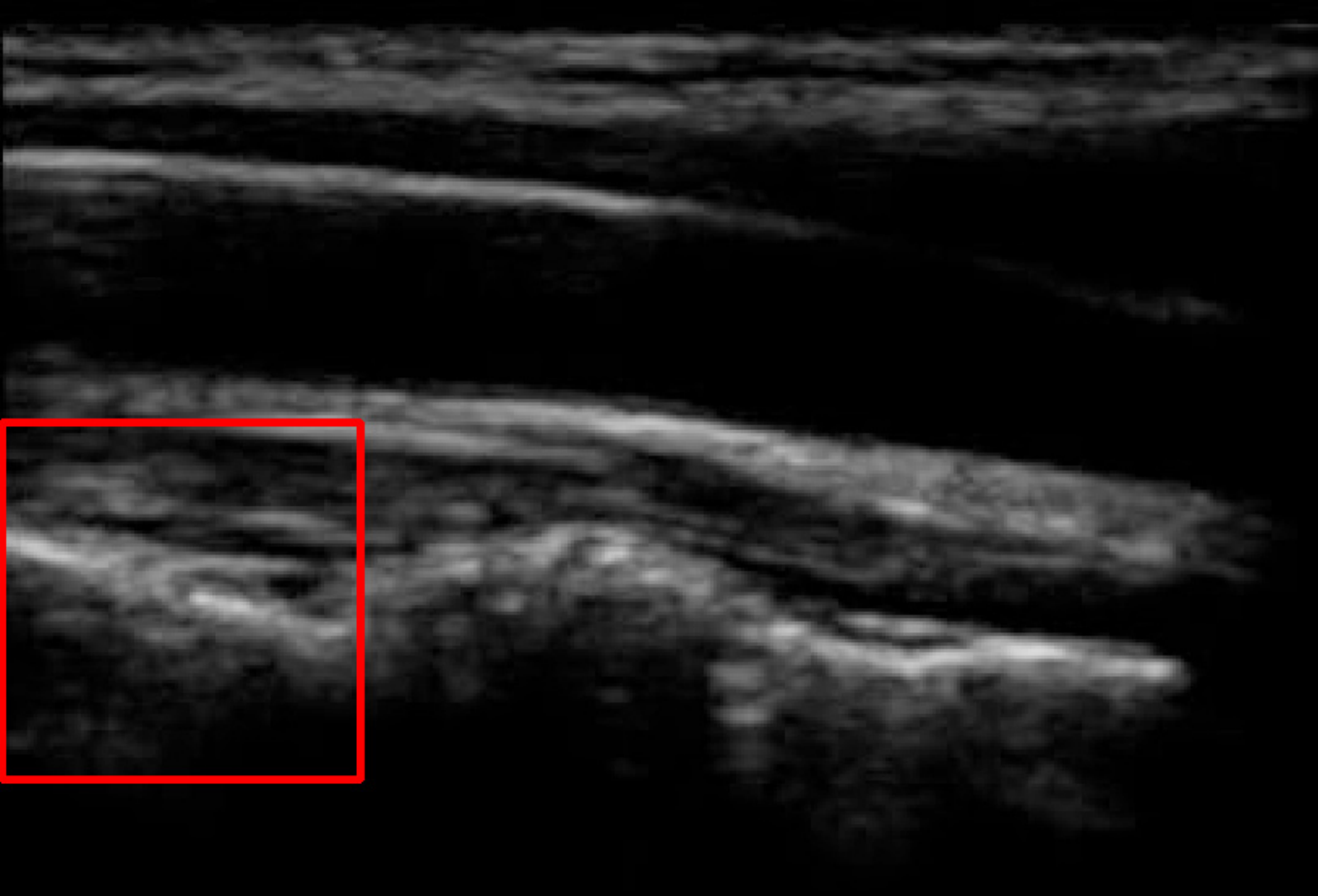}}
	\hspace{0.1ex} 
	\subfloat[HR] 
	{\includegraphics[height=1.0in,width=1.0in]{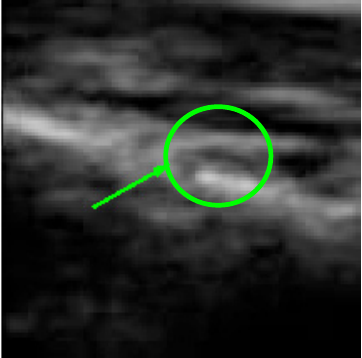}}
	\hspace{0.1ex} 
	\subfloat[SRCNN: \protect\\ 34.31/1.60] {\includegraphics[height=1.0in,width=1.0in]{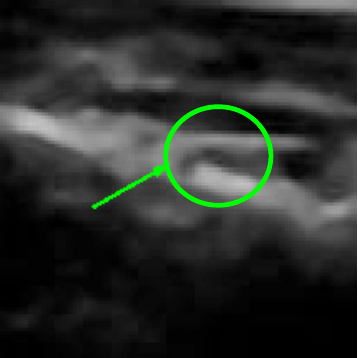}}
	\hspace{0.1ex} 
	\subfloat[SRGAN: \protect\\ 20.83/1.61]
	{\includegraphics[height=1.0in,width=1.0in]{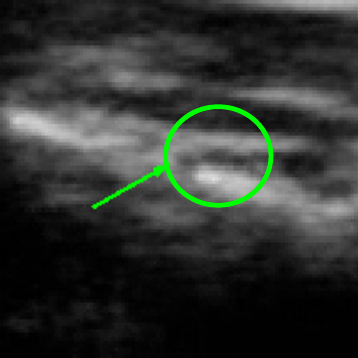}}
	\hspace{0.1ex}
	\subfloat[ZSSR: \protect\\ 35.31/2.25]
	{\includegraphics[height=1.0in,width=1.0in]{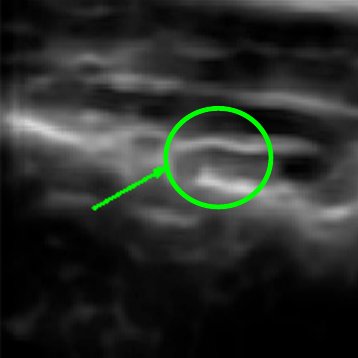}}
	\hspace{0.1ex} 
	\subfloat[The proposed method: \protect\\ 36.43/2.34]
	{\includegraphics[height=1.0in,width=1.0in]{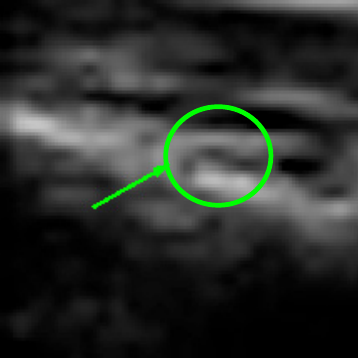}}
	\\
	\vspace{-2ex}
	\subfloat[Ground Truth]
	{\includegraphics[height=1.0in,width=1.3in]{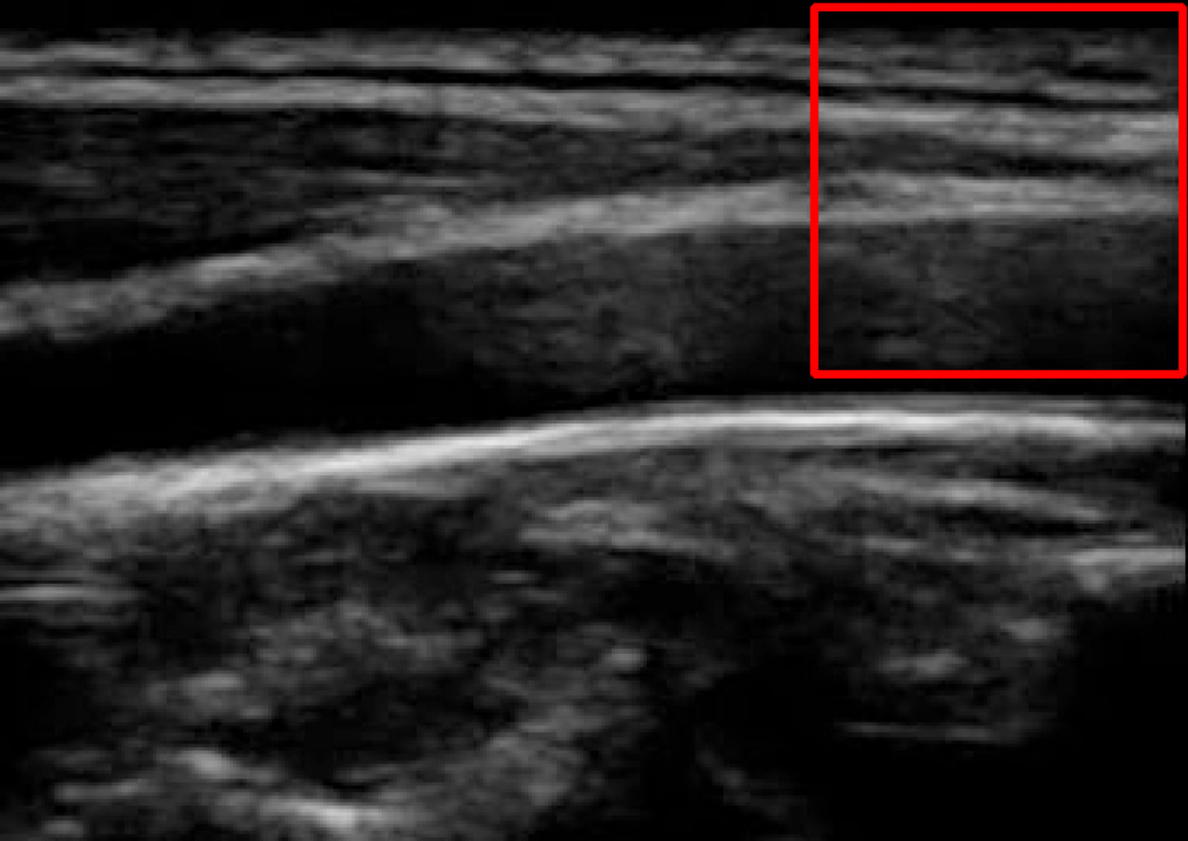}}
	\hspace{0.1ex} 
	\subfloat[HR]
	{\includegraphics[height=1.0in,width=1.0in]{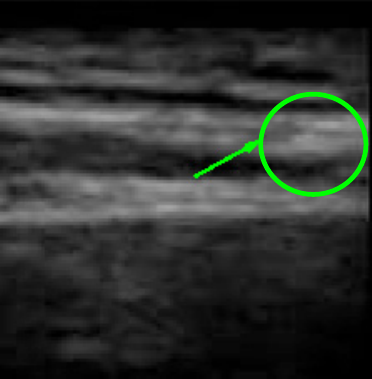}}
	\hspace{0.1ex} 
	\subfloat[SRCNN: \protect\\ 28.70/1.34]
	{\includegraphics[height=1.0in,width=1.0in]{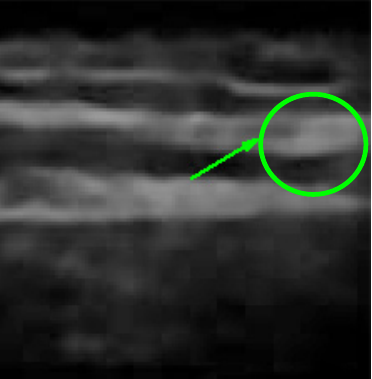}}
	\hspace{0.1ex} 
	\subfloat[SRGAN: \protect\\ 31.16/1.87]
	{\includegraphics[height=1.0in,width=1.0in]{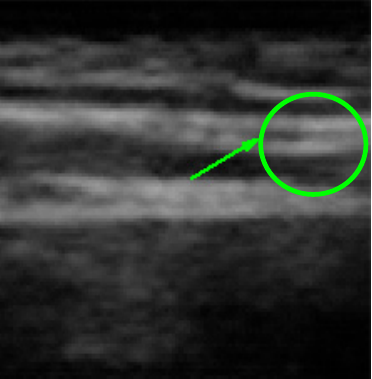}}
	\hspace{0.1ex}
	\subfloat[ZSSR: \protect\\ 30.17/2.312]
	{\includegraphics[height=1.0in,width=1.0in]{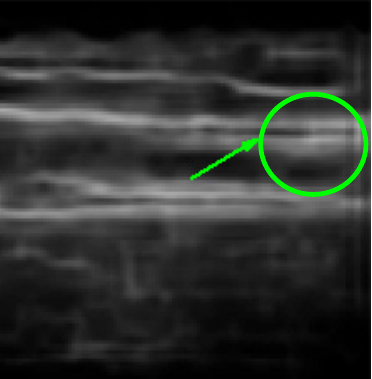}}
	\hspace{0.1ex} 
	\subfloat[The proposed method: \protect\\ 33.11/2.58]
	{\includegraphics[height=1.0in,width=1.0in]{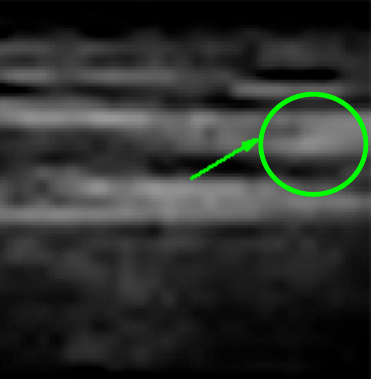}}
	\vspace{-2ex}
	\caption{The comparisons of visual effects and PSNR/IFC metrics for 4$\times$ super-resolved ultrasound images under CCA-US dataset by (b,h) Ground truth (c,i) SRCNN, (d,j) SRGAN, (e,k) ZSSR and (f,l) the proposed method. The green arrows and circles highlight the differences between the images}
	\label{fig5}
	\vspace{-2ex}
\end{figure*}

\begin{figure*}[!h]
	\captionsetup[subfigure]{justification=centering}
	\centering
	\subfloat[Ground Truth]
	{\includegraphics[height=1.0in,width=1.3in]{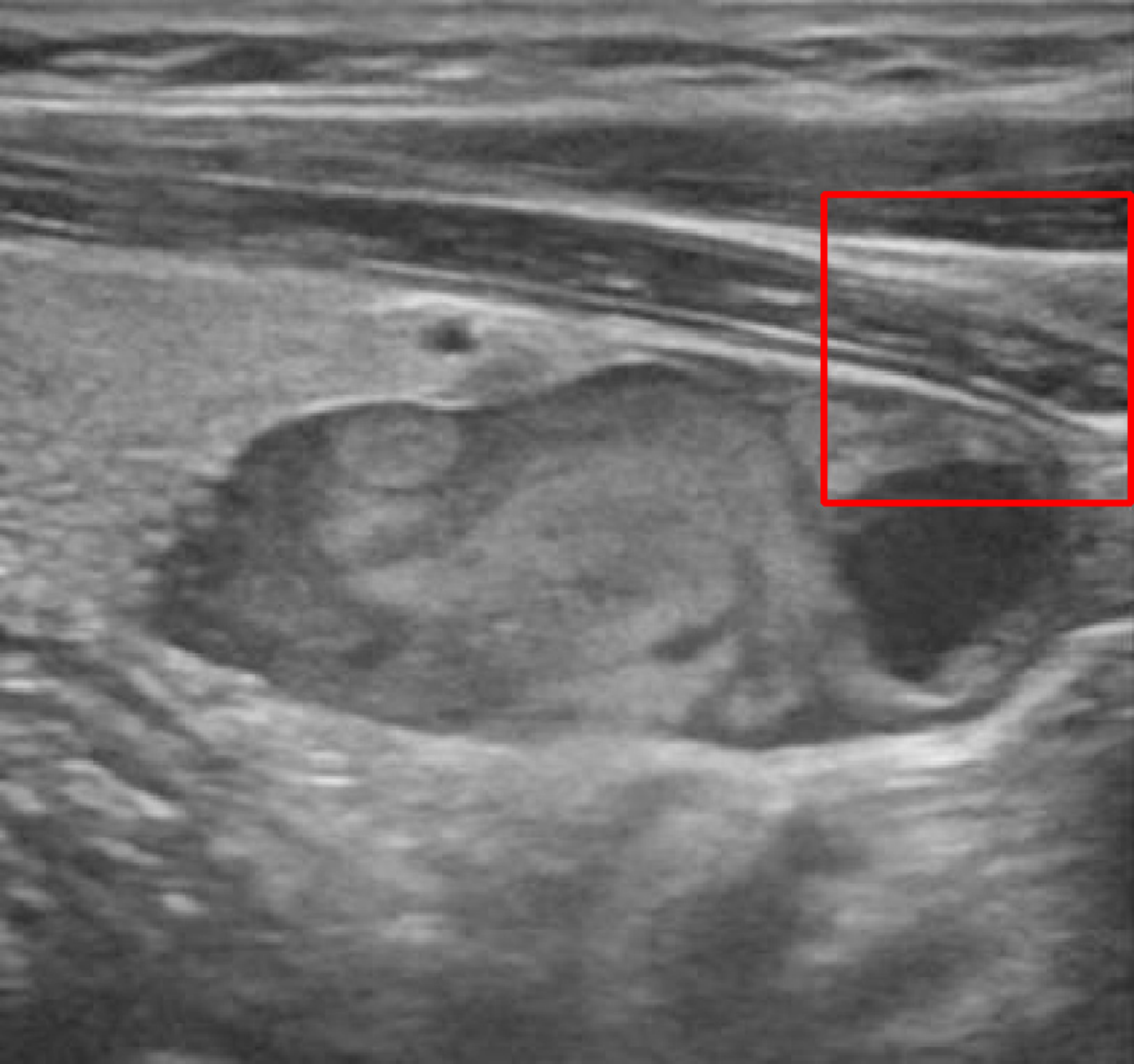}}
	\hspace{0.1ex} 
	\subfloat[HR]
	{\includegraphics[height=1.0in,width=1.0in]{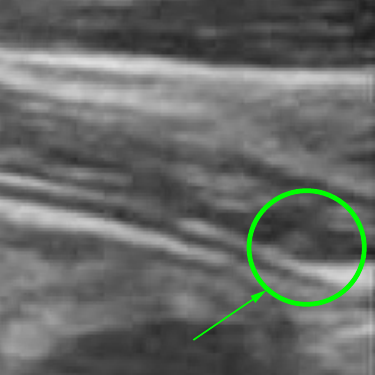}}
	\hspace{0.1ex} 
	\subfloat[SRCNN \protect\\ 26.89/1.90] {\includegraphics[height=1.0in,width=1.0in]{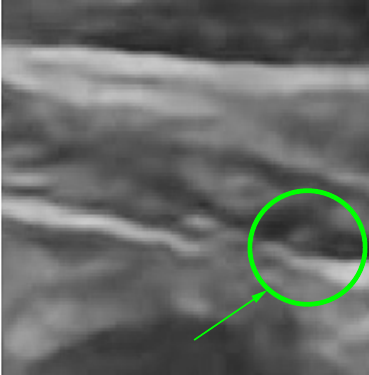}}
	\hspace{0.1ex}
	\subfloat[SRGAN \protect\\ 24.59/1.87] {\includegraphics[height=1.0in,width=1.0in]{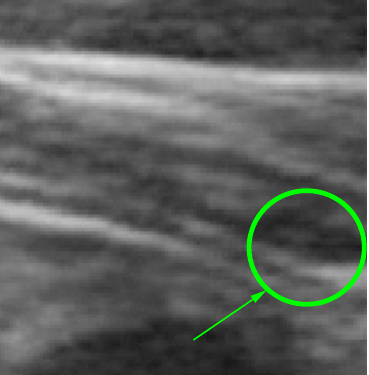}}
	\hspace{0.1ex}
	\subfloat[ZSSR \protect\\ 30.77/2.93]
	{\includegraphics[height=1.0in,width=1.0in]{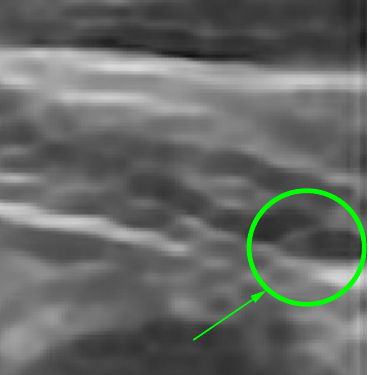}}
	\hspace{0.1ex} 
	\subfloat[The proposed method: \protect\\ 30.65/2.89]
	{\includegraphics[height=1.0in,width=1.0in]{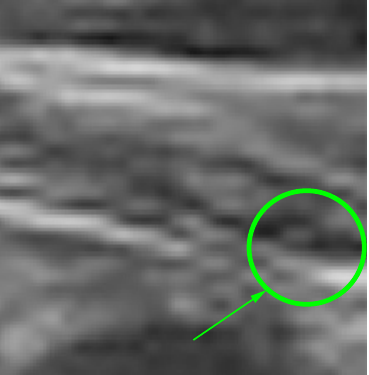}}
	\\
	\vspace{-1ex}
	\subfloat[Ground Truth]
	{\includegraphics[height=1.0in,width=1.3in]{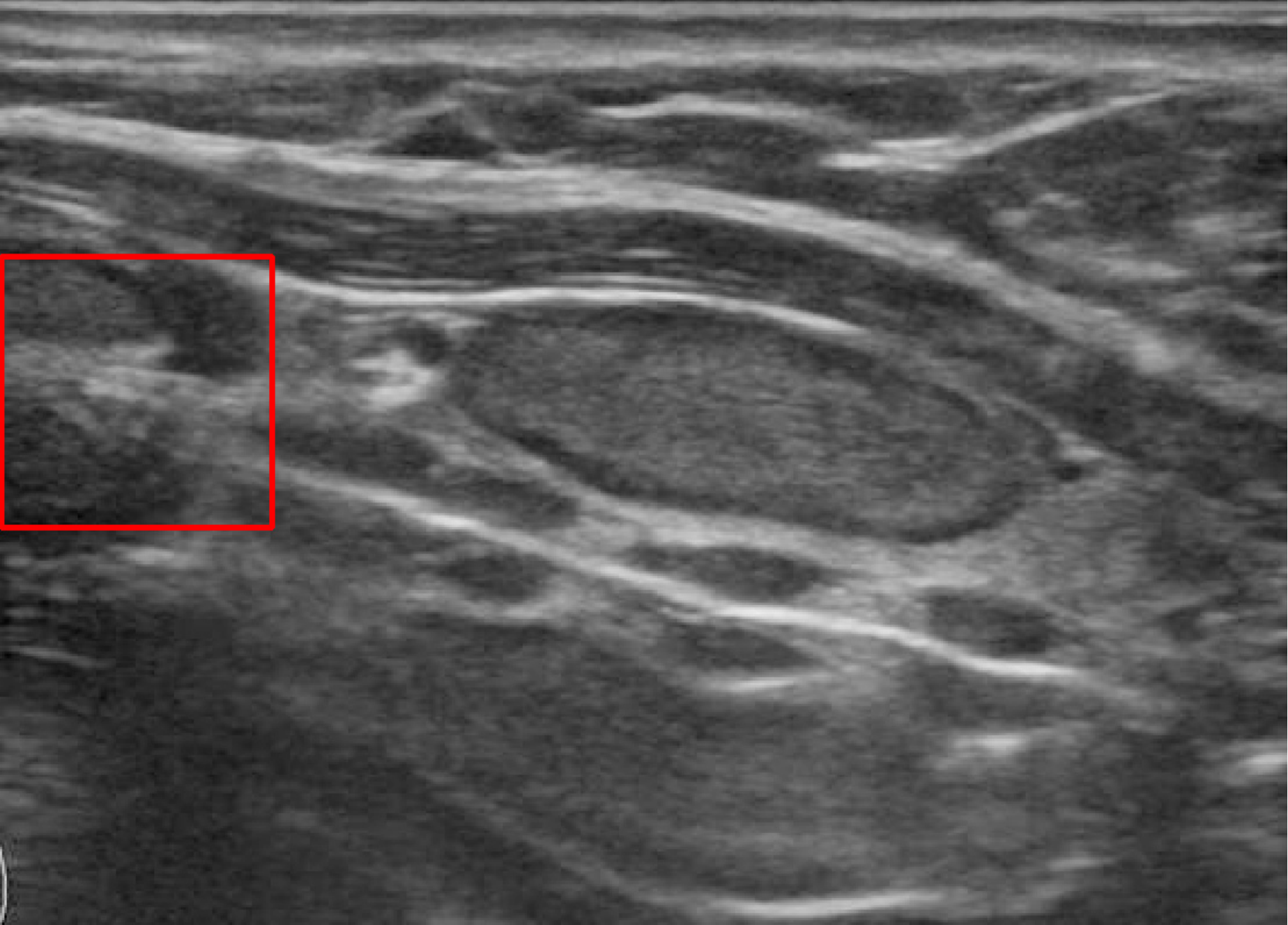}}
	\hspace{0.1ex} 
	\subfloat[HR]
	{\includegraphics[height=1.0in,width=1.0in]{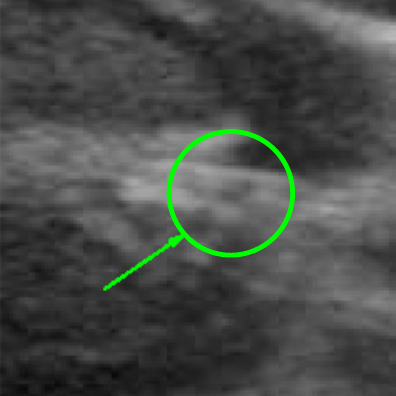}}
	\hspace{0.1ex} 
	\subfloat[SRCNN: \protect\\ 29.70/1.96] {\includegraphics[height=1.0in,width=1.0in]{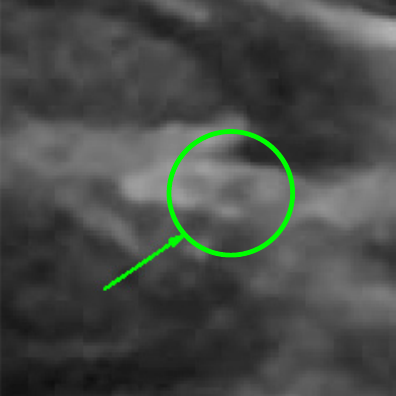}}
	\hspace{0.1ex}
	\subfloat[SRGAN: \protect\\ 25.98/2.02]
	{\includegraphics[height=1.0in,width=1.0in]{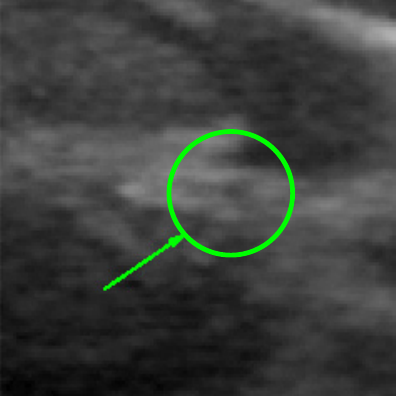}}
	\hspace{0.1ex} 
	\subfloat[ZSSR: \protect\\ 32.18/3.06]
	{\includegraphics[height=1.0in,width=1.0in]{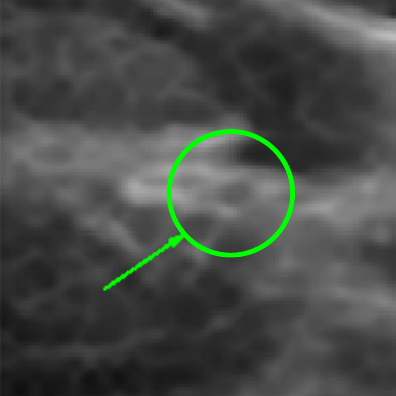}}
	\hspace{0.1ex} 
	\subfloat[The proposed method: \protect\\ 32.42/2.98]
	{\includegraphics[height=1.0in,width=1.0in]{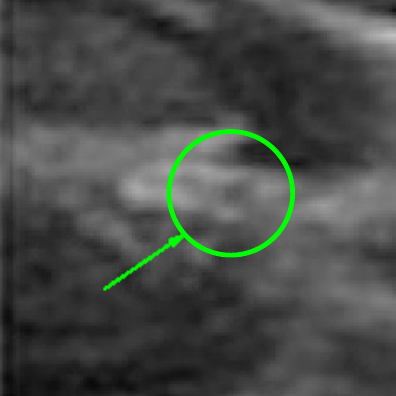}}
	\hspace{0.1ex} 
	\vspace{-2ex}
	\caption{The comparisons of visual effects and PSNR/IFC metrics for 4$\times$ super-resolved ultrasound images under US-CASE dataset by (b,h) Ground truth (c,i) SRCNN, (d,j) SRGAN, (e,k) ZSSR and (f,l) the proposed method. The green arrows and circles highlight the differences between the images}
	\label{fig6}
	\vspace{-2ex}
\end{figure*}

\begin{figure*}[t]
	\captionsetup[subfigure]{justification=centering}
	\centering
	\subfloat[Groud turth]
	{\includegraphics[height=1.4in,width=1.8in]{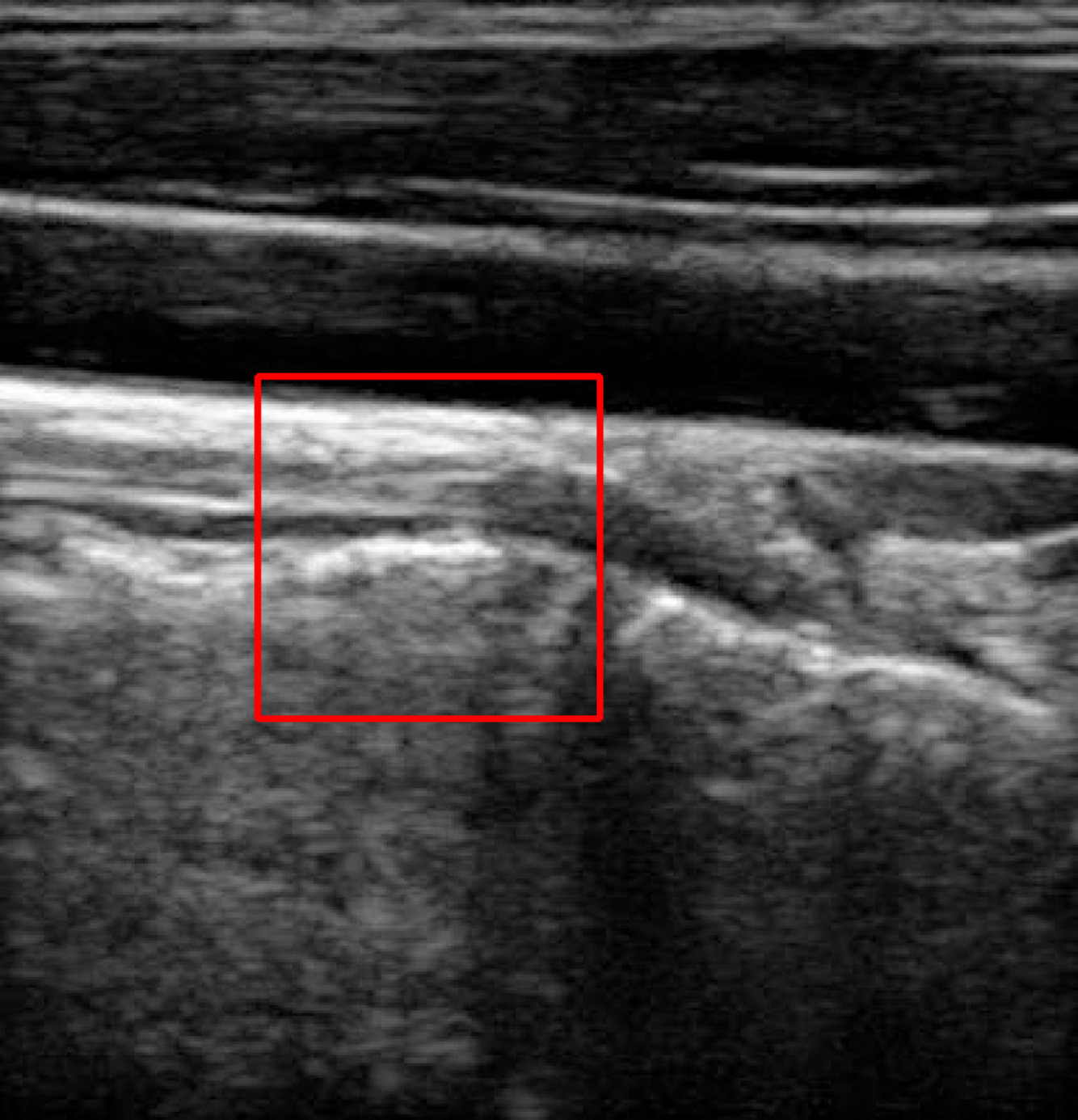}}
	\hspace{0.1ex} 
	\subfloat[HR]
	{\includegraphics[height=1.4in,width=1.4in]{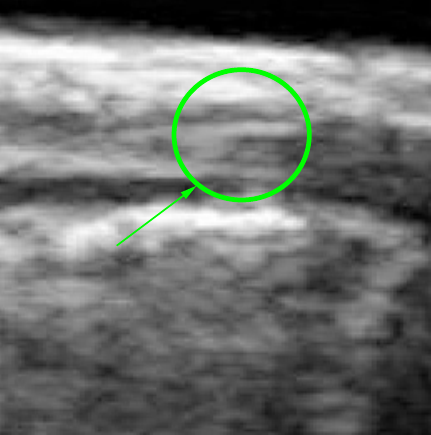}}
	\hspace{0.1ex} 
	\subfloat[ZSSR: \protect\\ 32.17/0.87] {\includegraphics[height=1.4in,width=1.4in]{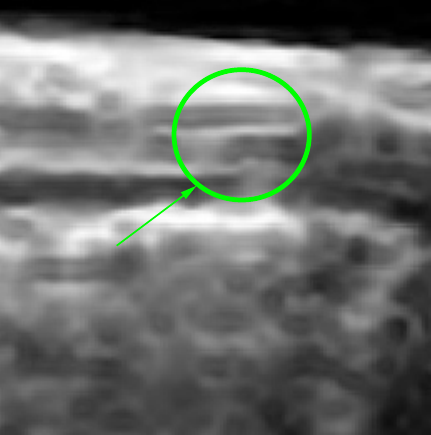}}
	\hspace{0.1ex} 
	\subfloat[Ours: \protect\\ 32.32/0.88]
	{\includegraphics[height=1.4in,width=1.4in]{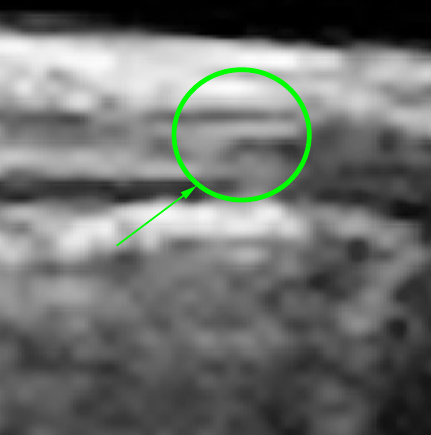}}
	
	\vspace{-2ex}
	\subfloat[Groud turth]
	{\includegraphics[height=1.4in,width=1.8in]{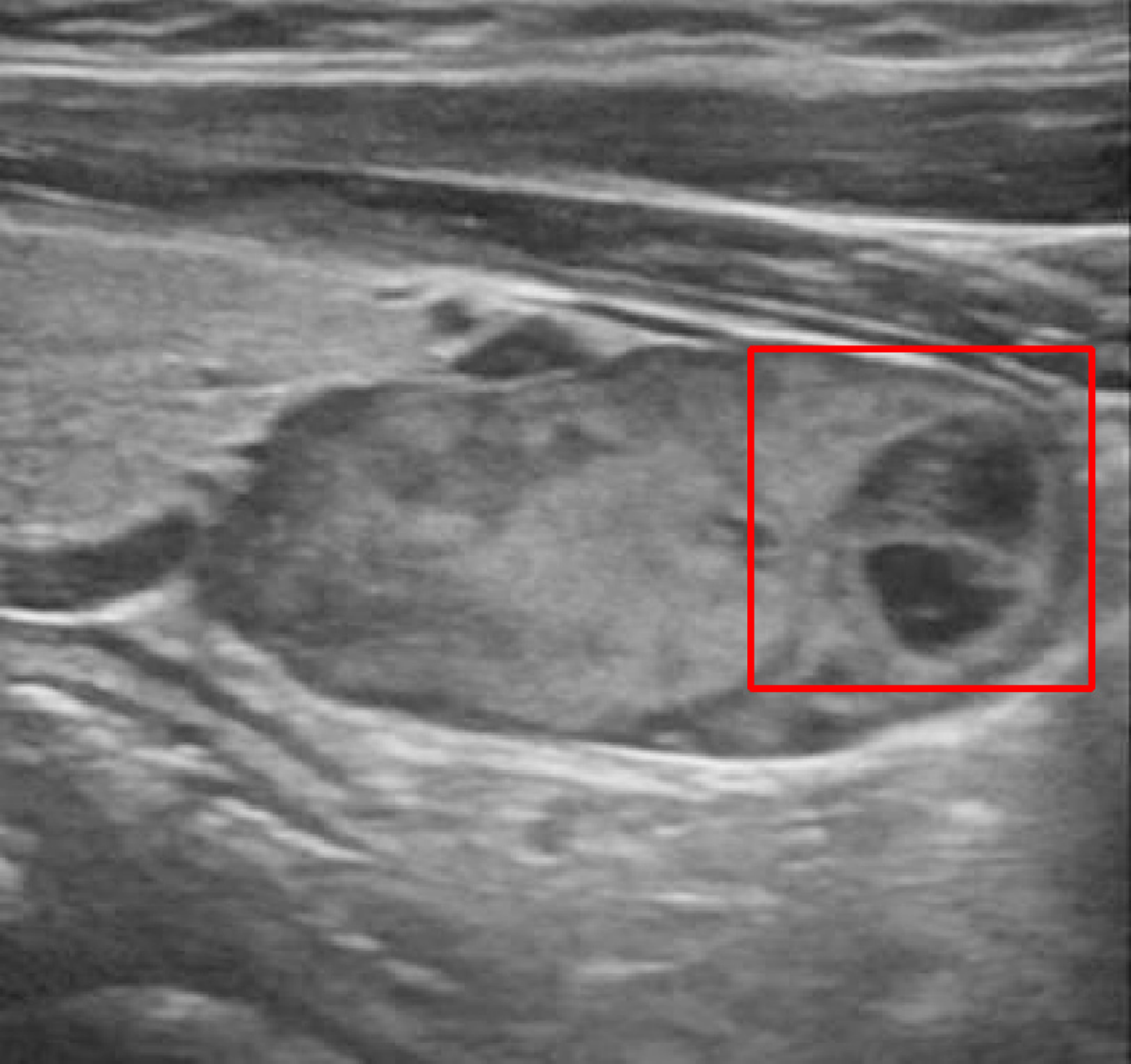}}
	\hspace{0.1ex} 
	\subfloat[HR]
	{\includegraphics[height=1.4in,width=1.4in]{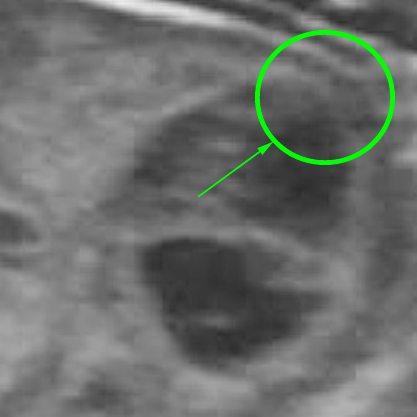}}
	\hspace{0.1ex} 
	\subfloat[ZSSR: \protect\\ 32.20/0.87] {\includegraphics[height=1.4in,width=1.4in]{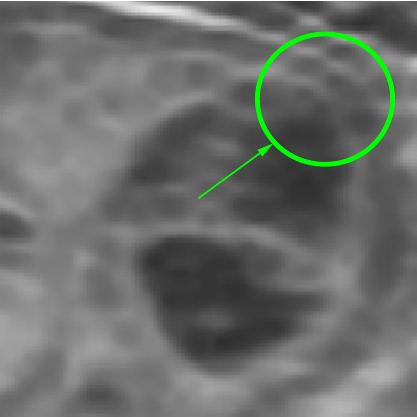}}
	\hspace{0.1ex} 
	\subfloat[Ours: \protect\\ 31.56/0.86]
	{\includegraphics[height=1.4in,width=1.4in]{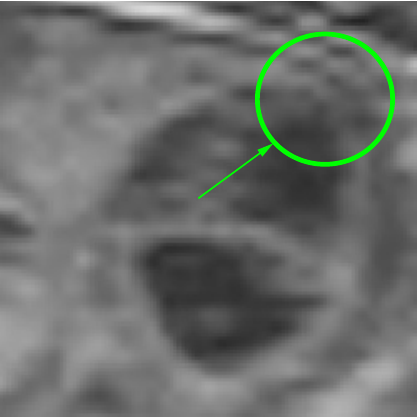}}
	\hspace{0.1ex}
	
	\caption{The comparisons of visual effects and PSNR/SSIM metrics with ZSSR. (a) from CCA-US dataset, (e) from US-CASE dataset. The green arrows and circles highlight the differences between the images}
	\label{fig7}
	\vspace{-2ex}
\end{figure*}

We provide some quantitative evaluation comparisons in Table \ref{table2} and Table \ref{table3}. In Fig. \ref{fig5},  Fig. \ref{fig6} and Fig. \ref{fig7},  we also provide some visual comparison examples. Moreover, in terms of running efficiency, we compare our approach with other methods in inference speed, model capacity, data processing throughput. The results are shown in Table \ref{table4}.
\begin{table}[t]
	\renewcommand{\arraystretch}{1.5}
	\caption{A comparison of PSNR and SSIM scores under two datasets. The bold numbers indicate the best results.}
	\label{table3}
	\renewcommand\tabcolsep{8.0pt}
	\centering
	\vspace{-1ex}
	\begin{threeparttable}[htpb]
		\begin{tabular}{l|l l l l}
			\hline
			\multirow{2}{*}{\textbf{DataSets}} & \multicolumn{2}{c}{US-CASE} & \multicolumn{2}{c}{CCA-US}						\\ \cline{2-5}
			& PSNR		& SSIM			& PSNR		& SSIM     \\
			\hline
			EDSR\cite{lim2017enhanced}                          & 25.290	& 0.740 & 27.432	& 0.804 \\
			
			SRFeat\cite{park2018srfeat}                            & 25.602 & 0.721 & 28.864 & 0.808 \\
			
			ZSSR\cite{ZSSR}                            			   & \textbf{32.670}  & 0.872  & 34.882 & 0.918	\\
			
			\hline
			Our proposed                          
			& 32.491 &  \textbf{0.876} & \textbf{35.222} & \textbf{0.919}    \\
			\hline
		\end{tabular}
	\end{threeparttable}
	\vspace{-3ex}
\end{table}
\begin{table*}[t]
	\renewcommand{\arraystretch}{1.5}
	\caption{Evaluation of the running efficiency for all methods. The bold numbers indicate the best}
	\label{table4}
	\centering
	\vspace{-1ex}
	\begin{threeparttable}[b]
		\begin{tabular}{l|l l l l l l}
			\hline
			& \textbf{SRCNN\cite{dong2016image}} & \textbf{SRGAN\cite{ledig2017photo}} & \textbf{SRFeat\cite{park2018srfeat}}&  \textbf{EDSR\cite{lim2017enhanced}}& \textbf{ZSSR\cite{ZSSR}} &\textbf{Our proposed}          \\
			\hline
			Platform & MATLAB & TensorFlow & TensorFlow & TensorFlow & pytorch & Pytorch \\
			Test Image Size & 600*488 & 150*112 & 150*112 & 150*112 & 150*112 & 600*448\\
			Inference Time & 188ms & \textbf{53ms} & 136ms & 49ms & 169ms & 176ms\\
			Throughput (Kb/ms) & 4.189 & 0.929 & 0.362 & 1.00 & 0.290 &\textbf{4.474}\\
			Model Capacity & \textbf{270KB} & 9.1MB & 37.2MB & 9.1MB & 3.6MB & 1.1MB        \\
			\hline
		\end{tabular}
	\end{threeparttable}
	\vspace{-2ex}
\end{table*}

Table \ref{table2} lists the comparison results of PSNR and IFC under a test data set consisting of 20 ultrasound images randomly selected from the two datasets mentioned above (each dataset selects 10 images). Compared with SRCNN \cite{dong2016image} and SRGAN \cite{ledig2017photo}, our method achieves the best results on both test images from CCA-US and US-CASE datasets. Table \ref{table3} lists the comparison results of PSNR and SSIM under the whole US-CASE and CCA-US datasets. We can see that our proposed method can attain the best or the second best PSNR results on the two ultrasonic datasets compared with EDSR \cite{lim2017enhanced}, SRFeat \cite{park2018srfeat}, ZSSR \cite{ZSSR}. AS for SSIM measures, our method will always achieve the best measurement results. On the whole, the performance of our method is better than others. In addition, the results in two tables suggest that the self-supervised learning methods (including ours and ZSSR) might have more prospects on SR task than those of supervised learning.

According to Fig. \ref{fig5} and Fig. \ref{fig6}, it is clear that comparing with other methods, our presented method acquires the better SR visual effects. Especially, observing the local details of these SR images in Fig. \ref{fig5} and Fig. \ref{fig6} carefully,  we can see that the results of our method are more accurate than others and do not introduce the artifacts or noise. In addition, Fig. \ref{fig7} shows additional visual details comparisons with ZSSR\cite{ZSSR}. From the figure, it is easy to find that, ZSSR  is likely to introduce some unwanted artifacts. For example in Fig. \ref{fig7} (a,b), there are always some artificial pore structure appeared in the generated images of ZSSR. These artifacts might cause misdiagnosis by clinicians. Our CycleGAN framework can effectively alleviate this issue to achieve relatively accurate visual effects although its PSNR/SSIM perhaps decrease slightly.

Furthermore, according to Table \ref{table4}, it is easy to find that the throughput of our proposed model achieve the best performance among all compared SR methods. This means that our model can concurrently process larger deal of image data than others. Moreover, from the table, it is clear the proposed model is a lightweight one due to the model capacity of ours only a little higher than the simplest model - SRCNN.

In general, our proposed method has good visual effects and preferable objective evaluation indicators, which is of great value for ultrasound visual diagnosis in the medical industry.
\subsection{Ablation Study}
In order to analyze the impact of the components on the loss function (Eq. \ref{eq10}) on ultrasound image SR performance, we develop some variants of our model: (1) GAN alone, where other losses including cycle loss are removed, (2) cycle alone , (3) GAN only with forward cycle (LR-SR-LR), (4) GAN only with backward cycle (HR-LR-SR) and (5) GAN with cycle. These variants are trained under the same condition as our original model. The results are presented in Table \ref{table5}.
\begin{table}[t]
	\renewcommand{\arraystretch}{1.5}
	\caption{Ablation study on CCA-US dataset. The best results are indicated in Bold.}
	\label{table5}
	\renewcommand\tabcolsep{10.0pt}
	\centering
	\vspace{-1ex}
	\begin{threeparttable}[]
		\begin{tabular}{l|l l}
			\hline
			\multirow{2}{*}{\textbf{DataSets}} &  \multicolumn{2}{c}{CCA-US}						\\ \cline{2-3}
			& PSNR		&IFC     \\
			\hline
			GAN alone                      & 33.968	& 2.203 \\
			Cycle alone					&{34.721} & 2.298	\\
			GAN + forward cycle			   & 34.282	& 2.221\\
			GAN + backward cycle            & 34.519  & 2.262      \\
			GAN + cycle						&	{34.839} & 	2.303  \\
			\hline
			Ours                          & \textbf{34.900} & \textbf{2.317}\\
			\hline
		\end{tabular}
	\end{threeparttable}
	\vspace{-2ex}
\end{table}

From Table \ref{table5}, it is obvious only using GAN (the adversarial loss), the performance of the results will be much reduced. While quite good performance can be achieved with only utilizing cycle loss. Meanwhile, the forward cycle loss and the backward cycle loss both contribute to the performance. The combination of cycle loss with GAN can achieve better results. Finally, all four losses proposed have an effect on the final reconstruction performance. Thus, we can conclude that the cycle structure is extremely beneficial to ultrasound image SR.


\section{Conclusion}
In this work, for medical industry, we propose a novel perception consistency ultrasound image SR approach based on self-supervised CycleGAN framework. Firstly, we analyze the multi-scale pattern characteristics between the local parts and the whole image for ultrasound data and propose to apply self-supervised learning strategy to get LR-HR pairs when lacking numerous ultrasound training images. Then we introduce a CycleGAN framework with a synthetic imaging loss, including the pixel-wise loss, the perceptual feature loss, the adversarial loss and the most important cycle consistency loss, to guarantee that the image ensemble and the details can keep the perception consistency not only in LR-to-SR-to-LR cycle but also in HR-to-LR-to-SR one. According to the evaluation results under two ultrasound datasets, it is clearly demonstrated that the proposed self-supervised  CycleGAN approach achieves the best performance not only in objective qualitative results and the running efficiency but also in visual effects.

In the meantime, it should be noted that ultrasound data SR may pay more attention to the accuracy of reconstruction than that of natural images. Therefore, our near future work will center on extending the proposed approach to natural image tasks, such as background subtraction \cite{sakkos2018end}, image defogging \cite{liu2017large}, etc., and analyzing the relationship between reconstruction accuracy and visual effects.


%



\section*{Acknowledgment}
We thank all the students in our Lab of AHUT for their help in discussions. This work was supported in part by the National Natural Science Foundation of China under Grant No. 61971004, the Natural Science Foundation of Anhui Province under Grant No. 2008085MF190 and Grant No. 1808085QF210, and also by the Key Project of Natural Science of Anhui Provincial Department of Education under Grant No. KJ2019A0083.

\ifCLASSOPTIONcaptionsoff
  \newpage
\fi



%
\bibliographystyle{./IEEEtran}

%








\end{document}